\documentclass[aps, prd, twocolumn, superscriptaddress, footinbib]{revtex4}
\usepackage{graphicx}
\usepackage{dcolumn}
\usepackage{amssymb}
\usepackage{amsmath,amssymb,amsfonts}

\DeclareMathOperator{\sinc}{sinc}

\begin{document}


\newcommand{\vev}[1]{\langle #1\rangle}
\newcommand{\Eq}[1]{\mbox{Eq. (\ref{eqn:#1})}}
\newcommand{\Fig}[1]{\mbox{Fig. \ref{fig:#1}}}
\newcommand{\Sec}[1]{\mbox{Sec. \ref{sec:#1}}}

\newcommand{\PHI}{\phi}
\newcommand{\PhiN}{\Phi^{\mathrm{N}}}
\newcommand{\vect}[1]{\mathbf{#1}}
\newcommand{\Del}{\nabla}
\newcommand{\unit}[1]{\;\mathrm{#1}}
\newcommand{\x}{\vect{x}}
\newcommand{\y}{\vect{y}}
\newcommand{\p}{\vect{p}}
\newcommand{\ScS}{\scriptstyle}
\newcommand{\ScScS}{\scriptscriptstyle}
\newcommand{\xplus}[1]{\vect{x}\!\ScScS{+}\!\ScS\vect{#1}}
\newcommand{\xminus}[1]{\vect{x}\!\ScScS{-}\!\ScS\vect{#1}}
\newcommand{\diff}{\mathrm{d}}
\newcommand{\mk}{{\mathbf k}}
\newcommand{\ep}{\epsilon}

\newcommand{\be}{\begin{equation}}
\newcommand{\ee}{\end{equation}}
\newcommand{\bea}{\begin{eqnarray}}
\newcommand{\eea}{\end{eqnarray}}
\newcommand{\vu}{{\mathbf u}}
\newcommand{\ve}{{\mathbf e}}
\newcommand{\vn}{{\mathbf n}}
\newcommand{\vk}{{\mathbf k}}
\newcommand{\vhk}{{\mathbf {\hat k}}}
\newcommand{\vb}{{\mathbf b}}
\newcommand{\vhb}{{\mathbf {\hat b}}}
\newcommand{\vz}{{\mathbf z}}
\newcommand{\vx}{{\mathbf x}}
\newcommand{\vy}{{\mathbf y}}
\def\dup{\;\raise1.0pt\hbox{$'$}\hskip-6pt\partial\;}
\def\ddn{\;\overline{\raise1.0pt\hbox{$'$}\hskip-6pt\partial}\;}

\title{Unsqueezing of standing waves due to inflationary domain structure}

\newcommand{\addressImperial}{Theoretical Physics, Blackett Laboratory, Imperial College, London, SW7 2AZ, United Kingdom}
\newcommand{\addressRoma}{Dipartimento di Fisica, Universit\`a ‚ÄúLa Sapienza‚Äù
and Sez. Roma1 INFN, P.le A. Moro 2, 00185 Roma, Italia}

\author{Carlo~R.~Contaldi}
\email{c.contaldi@imperial.ac.uk}
\author{Jo\~{a}o Magueijo}
\affiliation{\addressImperial}

\date{\today}

\begin{abstract}
The so-called ``trans-Planckian'' problem of inflation may be evaded by positing that modes come into existence only when they became ``cis-Planckian'' by virtue of expansion. However, this would imply that for any mode a new random realization  would have to be drawn 
every $N$ wavelengths, with $N$ typically of order 1000 (but it could be larger or smaller). Such a re-drawing of realizations leads to a heteroskodastic distribution if the region under observation contains several such independent domains. This has no effect on the sampled power spectrum for a scale-invariant raw spectrum, but at very small scales it leads to a spectral index bias towards scale-invariance and smooths oscillations in the spectrum. The domain structure would also ``unsqueeze'' some of the propagating waves, i.e., dismantle their standing wave character. By describing standing waves as travelling waves of the same amplitude moving in opposite directions we determine the observational effects of unsqueezing. We find that it would erase the Doppler peaks in the CMB, but only on very small angular scales, where the primordial signal may not be readily accessible. The standing waves in a primordial gravitational wave background would also be turned into travelling waves. This unsqueezing of the gravitational wave background may constitute a detectable phenomenon.  
\end{abstract}

\keywords{cosmology, gravitational waves, cosmic microwave background}
\pacs{}

\maketitle
\section{Introduction}

Late--time temporal coherence of cosmological density fluctuations underpins the existence of ``Doppler peaks'' in the Cosmic Microwave Background (CMB). Indeed the presence of peaks in the CMB angular power spectrum is one of the most profound discoveries made in over three decades of observing CMB anisotropies~\cite{Netterfield:2001yq}. Their presence has been used to argue that perturbations driving acoustic fluctuations in the pre--recombination, tightly--coupled baryon--photon fluid must have re--entered the horizon after a period where they were driven to super--horizon scales by an epoch of accelerating expansion. The peaks therefore support inflationary-type scenarios for the origin of primordial perturbations, as opposed to {\it active} mechanisms operating on sub--horizon scales~\cite{active-passive}. Nonetheless we should add that, restricting to passive scenarios, inflation is not the only scenario to ``squeeze'' the fluctuations~\cite{squeeze,squeeze1}. Indeed it is hard to find a passive scenario that does not squeeze or has the same practical effects of squeezing. 

According to the inflationary paradigm the density fluctuations, as well as the primordial gravitational wave background (should there be one), were once micro--physical vacuum {\it quantum} fluctuations, which were then blown out of the Hubble radius by  inflationary expansion. At some point in their life these quantum fluctuations became {\it classical}, in a process still shrouded in mystery~\cite{starob95,starob96,pedro,martin, Martin:2012pea,Kiefer:1998qe, Kiefer:2008ku, Perez:2005gh,Nelson:2016kjm,Burgess:2014eoa,drubo}. 
Several explanations for the replacement of quantum uncertainty with classical probability have been put forward, with ``squeezing'' after first Hubble crossing usually playing a role~\cite{pedro,starob95,starob96}. 
The gravitational interaction responsible for squeezing can act as the ``observer''~\cite{Nelson:2016kjm,Burgess:2014eoa}, but other types of interactions may have caused the collapse of the wave--function~\cite{drubo}. 

Independently of the exact resolution of this problem, 
in this paper we focus on a closely related matter. Given the statistical nature of the cosmological perturbations, it is usually assumed that we observe a realization of the cosmological fluctuations drawn from a probability distribution (a process sometimes labelled ``cosmic variance''). The folklore is that ``observers in distant parts of the Universe'' would see a different realization. However, it is never specified how far apart these observers should be in order to see an independent random draw. 
It could even be that for some modes we probe different draws within our current Hubble volume. In this paper we relate this important issue with the problem of the collapse of the wave--function of the inflationary fluctuations. The trans-Planckian problem will appear as a crucial element in the discussion, and we start by briefly posing it~\cite{trans-Martin,trans-Staro,trans-Kofman}. 

In inflationary scenarios the ratio between the Hubble and the Planck lengths is a number $N=1/(H L_{\rm Pl})$ typically of order 1000. 
This implies that in just a few $e$-foldings (around 7) the modes evolve from being Planck scale to leaving the Hubble scale.
But, even in the most restrained models, inflation was in action for many more $e$-foldings before the scales we observe today left the Hubble radius. In fact, this is a requirement for solving a number of cosmological problems. Therefore, the modes we observe nowadays must have been nominally trans-Planckian in the early stages of inflation, i.e. had higher frequency than the Planck energy. This raises questions on the nature of the modes before they were stretched out enough to become cis-Planckian (assuming such stretching applied to them in that phase at all). Did the unknown quantum gravity physics ruling that regime affect the end-product of the inflationary mechanism of structure formation?

One possible way around the problem is to argue that we should not even discuss this question. Let us first ignore expansion and assume a quantum gravity scenario in which space-time is quantum and foam-like on length scales smaller than the Planck length, so that the concepts of fixed metric and locality do not make sense. 
In such a picture, when we consider QFT in Minkowski space-time we simply refrain from discussing trans-Planckian modes. In the same vein, in an expanding universe we should not discuss fixed comoving-length modes {\it before} their nominal physical length becomes cis-Planckian. The two situations are not precisely equivalent: the difference is that in the case of an expanding universe we must then account for the modes creation in time, in a process that is necessarily non-unitary. We shall not depend on the details of such a process in the discussion that follows, but suffice it to say that consistent non-unitary formalisms have been proposed allowing for such a process. 

\begin{figure}[t]
\centering
\includegraphics[trim={3cm 8cm 1cm 6cm},clip, width=9cm,angle=0]{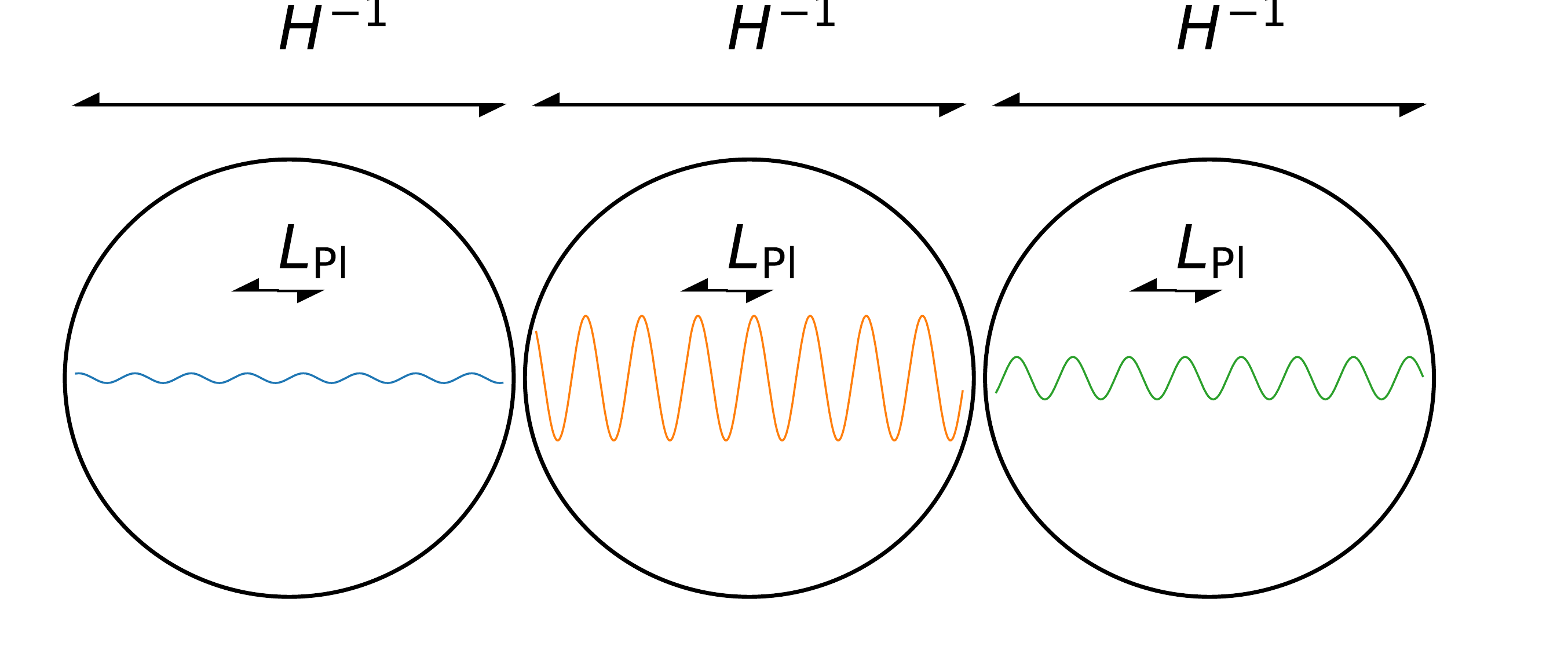}
\caption{A schematic representation of our domain setup. In each region, initially separated by a Hubble $H^{-1}$ length, quantum fluctuations are created at the Planck scale $L_{\rm Pl}$ that is $N$ times smaller than the Hubble scale. Consequently, the modes we see today, stretching over many inflationary Hubble scales, have a domain structure whereby each $\sim N\lambda$ sized patch has independent amplitudes and phases drawn from the same Gaussian distribution.}\label{fig:domains}
\end{figure}

In this paper we show that, while this evasive tactic
might be justified, 
its ultimate implications would be no less dramatic. In such a picture, 
the conventional vacuum state for a given mode is set up only when its physical wavelength reaches the Planck length, but because of causality it could not spatially extend beyond the Hubble volume, and thus not apply to a packet larger than 
$N\sim 1000$ wavelengths across. The overall vacuum state would therefore be the tensor product of such vacuum states applicable to independent domains. As soon as the state for each domain is prepared, expansion starts causally disconnecting its various parts;
however, they will remain entangled until collapse. When the wavelength reaches the Hubble scale and collapse occurs, these $N^3$ entangled regions will collapse into the same realization. Should the collapse take place before, or after Hubble first crossing, the picture does not qualitatively change (although the value of $N$ will do, inducing a quantitative dependence on the exact collapse mechanism). Regardless of the details,  
we will always have a situation in which a random draw is taken every $N$ wavelengths, for some $N$, creating a patchwork of independent ``domains''. This setup is sketched in Fig.~\ref{fig:domains} for a single mode. The rest of the paper is devoted to exploring the observational consequences of this fact.

In Section~\ref{sec:power} we first consider the case where the original fluctuations in each domain remain there (either because they are still
frozen-in or because they have zero momentum) and we sample their power spectrum using a region containing several such domains. This situation is only observationally relevant for measurements of dark matter fluctuations. For radiation and gravity waves evolution after horizon re-entry is inevitable, a matter studied in subsequent Sections.
Then, we find that if more than one domain were present in the field, there would be a strong dilution effect in the sampled power spectrum, should the raw power spectrum be a delta function in wavenumber. For more realistic spectra, leakage from other modes makes up for this dilution; indeed, for a scale invariant power spectrum, sampled and raw spectrum are the same. For spectra with features these would be smoothed out, and for tilted spectra the sampled spectrum would also be different from the raw one. In general the process would be heteroskodastic, and be non-Gaussian, resulting from the marginalization of a Gaussian distribution with random parameters.

In Section~\ref{sec:unsqueeze} we address in more detail the impact upon the CMB radiation. 
We review the difference between travelling and standing waves and how it relates to the squeezing of modes during the inflating phase. 
We argue that the structure of domains leads to ``unsqueezing'' of modes. Standing waves can be seen as travelling waves with the same real amplitude moving in opposite directions. As the waves move away from their original domains they are faced with waves travelling in opposite directions which originated in different domains, and so, have a different amplitude. An ensemble of travelling waves is thus produced out of originally standing waves. 
This has repercussions on the presence of Doppler peaks on very small scales.

In Section~\ref{sec:gws} we focus instead on the primordial Gravitational Wave Background (GWB). In this case, we argue that unsqueezing of modes is, in principle, directly observable using gravitational wave interferometry. This occurs because of the ability of gravitational wave detectors to interfere waves travelling with {\it opposite} momenta \footnote{This is distinct from radio interferometry where the presence of a compact beam on the sky means waves with opposite momentum cannot be interfered.}. The presence, or lack of, standing waves in this measurement leads to a typical modulation of the short--term interference signal, in analogy to a well--known observational effect in radio astronomy.

We conclude in Section~\ref{sec:discussion} with a discussion of our main results.

\section{Power spectrum of a structure of domains}\label{sec:power}
Let us first assume that the fluctuations imparted upon each domain 
do not move away from it until they are sampled by a late time observer. We are interested in the power spectrum that would be measured should we sample the fluctuations with a region (field) containing several domains. In practice, the assumption made will be a bad one except for the dark matter distribution (assuming zero speed of sound) and other fluctuations with negligible momentum. Nonetheless, some of the results presented in this Section will be relevant in latter Sections.  We refer the reader to Appendix~\ref{app:domains} for all the details in this Section. For simplicity we illustrate our points in one dimension, but the results presented generalize easily to three dimensions. 

Let us consider a situation in which for each Fourier mode a different random draw of amplitude and phase is taken from the same Gaussian distribution every $N$ wavelengths. Let us assume we observe the phenomenon from a box of comoving size $L$, so that the (comoving) momenta are 
\begin{equation}
k_n = \frac{2 \pi}{L} n  \;\;\Rightarrow\;\; \lambda_n = \frac{L}{n} \, .
\end{equation} 
The fluctuations' field (here generically denoted by $\zeta$) can then be written as:
\begin{equation}
\zeta \left( x \right) = \sum_{n=-\infty}^{\infty} \sum_{j=1}^{J_n} \zeta_{n,j} \, e^{i k_n x} \, W_{n,j} \left( x \right) \,,
\end{equation} 
where $\zeta_{-n,j}=\zeta^\star_{n,j}$.
The window function $W_{n,j}$ (with $W_{-n,j}=W_{nj}$), encoding the transition between domains, can in principle be anything (and the transition be smooth or not). A simple choice is
\begin{equation}
W_{n,j} \left( x \right) = \left\{ \begin{array}{l} 
1 \;\;\; {\rm if } \;\; x_{n,j} < x < x_{n,j+1} \;\; {\rm and } \;\; 0 \le x \le L \\ 
0 \;\;\; {\rm otherwise} 
\end{array} \right.\,,
\end{equation} 
with 
\begin{equation} 
x_{n,j} \equiv x_{0n} + \left( j - 1 \right) N \, \lambda_n \,.
\end{equation} 
Note that the domains have different sizes (typically $\ell_{n,j}\sim N\lambda_n$) for different wavelengths, $\lambda_n$. They may also be shifted by a random $x_{0n}$ independently for different modes. This distinguishes our situation from more conventional windows.
We assume:
\begin{equation} \label{corr-ni}
\left\langle \zeta_{n,i} \zeta^*_{m,j} \right\rangle = \delta_{ij} \delta_{nm} \, P \left( k_n \right)\,,
\end{equation} 
where $P(k)$ is the raw power spectrum predicted by the inflationary model. 

In Appendix~\ref{app:domains} we evaluate in detail the power spectrum ${\hat P}(k)$ that would be sampled using a box of size $L$, given such a domain structure. Is this appreciably different from the raw $P(k)$? Given the unusual nature of our window (it is wavelength dependent), it turns out that the answer depends crucially on which scale we are measuring. For a box with size $L$ the ``turning-point'' is the scale: 
\begin{equation}
k_N=\frac{2\pi}{\lambda_N}=\frac{2\pi}{L}N\,.
\end{equation}
For this mode $\ell_N=N\lambda_N=L$, and so its associated domain size equals the box size. For $k\gg k_N$ there are many independent domains in the field, whereas for $k\ll k_N$ only one domain will typically be sampled (the smaller the $k$ the more unlikely it is to find a transition between domains within the field). Around $k\sim k_N$, we sample at most a couple of partial domains, and the sampled spectrum will depend on where the transition is located within the box. This renders the calculation more intricate. 

It is important not to let $L\rightarrow \infty$ seeking formal simplification. The size of the observation box  matters. For a given wavelength $\lambda$ it makes a world of difference whether we probe it with a box much smaller or much larger than $N\lambda$. Reciprocally, for a given observation box, it makes a lot of difference whether we are in the $k\gg k_N$ regime or in the $k\ll k_N$, or in the transition between the two.  

\subsection{The power spectrum in the transition regime $k\sim k_N$}
It is obvious that for $k\ll k_N$ we have $\hat P(k)\approx P(k)$, but the situation is a lot more involved in the transition region $k\sim k_N$. As shown in Appendix~\ref{app:domains} if leakage could be ignored (which only happens when the raw spectrum is a delta function and we focus observations on that mode), then the observed spectrum becomes dependent on $x_{0n}$, since this determines how many domains there are in the field, and their sizes. In this case, we must distinguish between the power spectrum conditional to a given set of domain phases $\{x_{0n}\}$, and the power spectrum marginalized over the distribution of $\{x_{0n}\}$. We label the conditional power spectrum $\hat P(k |x_{0n})$. The power spectrum marginalized over the $\{x_{0n}\}$ is then defined as:
\begin{equation}
\tilde P=\int dx_{n0} \, {\cal P}(x_{n0}) \hat P(k |x_{n0})\,. 
\end{equation}
In Appendix~\ref{app:domains} we find (ignoring leakage between modes) that:
\begin{equation}
\tilde P(k) = \left\{ \begin{array}{lr}P(k)\left(1-\frac{1}{3}\frac{k}{k_N}\right) &  {\rm for}\quad  k\le k_N\,,\\ \\P(k)\left(\frac{k_N}{k}-\frac{1}{3}\left(\frac{k_N}{k}\right)^2\right) & {\rm for}\quad  k\ge k_N\,,\end{array}\right.
\end{equation}
with more complex expressions for $\hat P(k |x_{n0})$ (cf. (\ref{Pkcond1}), (\ref{two_domain}), (\ref{Pkcond3}) and (\ref{Pkconf4})). The fact that these two are different signals an oddity that disappears in the limit $k\gg k_N$.

In fact we have introduced a heteroskodastic process, which may be regarded as a Gaussian or non-Gaussian one depending on how we set up the ensemble. 
For a given set of $\{x_{0n}\}$ the distribution is still Gaussian, but its correlation breaks homogeneity, since it depends on the $\{x_{0n}\}$, and therefore on the places where the transitions between domains happen (something signalled by the fact that the correlation has off-diagonal elements, had we computed them). However, a larger ensemble may be generated by the concatenation of these sub-ensembles, or seen another way, the full probability could be obtained by marginalizing the Gaussian distribution dependent on the $\{x_{0n}\}$ over these parameters. The full ensemble therefore would then be non-Gaussian but homogeneous. Its correlation is diagonal, but the higher order correlators are non-trivial. 

We defer to a future publication the study of this phenomenon taking leakage into account. As the next subsection shows, this can never be neglected for realistic power spectra. Nonetheless, even with the imperfect approximation used here, we were able to illustrate an important feature in the transition regime $k\sim k_N$.

\begin{figure}
\centering
\includegraphics[
width=3.5in,angle=0]{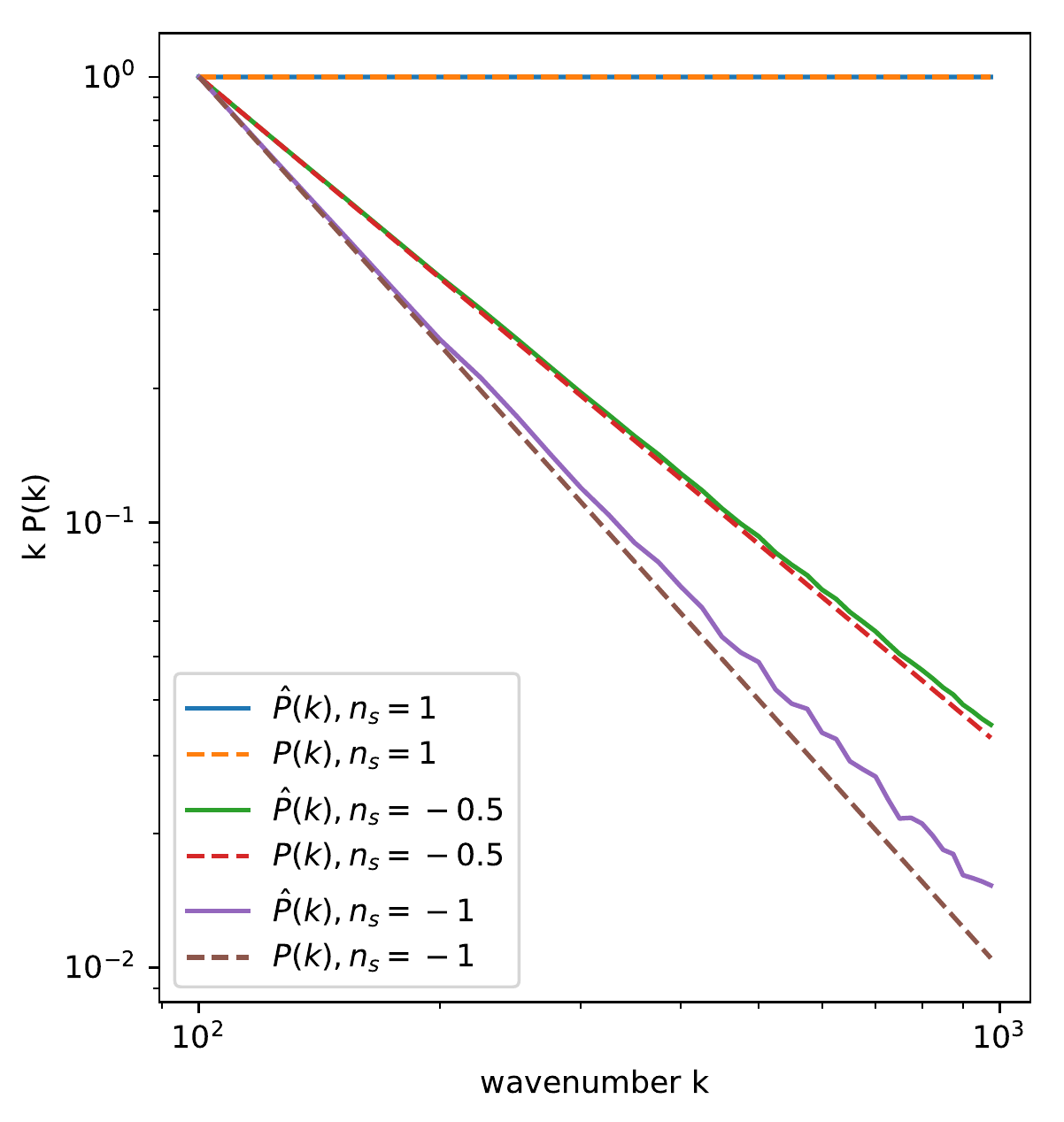}
\caption{An example of the differences between $P(k)$ (solid) and $\hat P(k)$ (dashed) for spectra with different tilts: $n_s=1$, $n_s=-0.5$ and $n_s=-1$. For illustrative purposes we have set $N=10$. We see that the more the spectrum is tilted the stronger the bias towards smaller tilt in $\hat P$ due to smoothing. The small scale structure in $\hat P$ is due to the aliasing of modes because of the finite sampling of the underlying, windowed spectrum.} \label{fig:tilted}
\end{figure}

\begin{figure}
\centering
\includegraphics[ 
width=3.5in,angle=0]{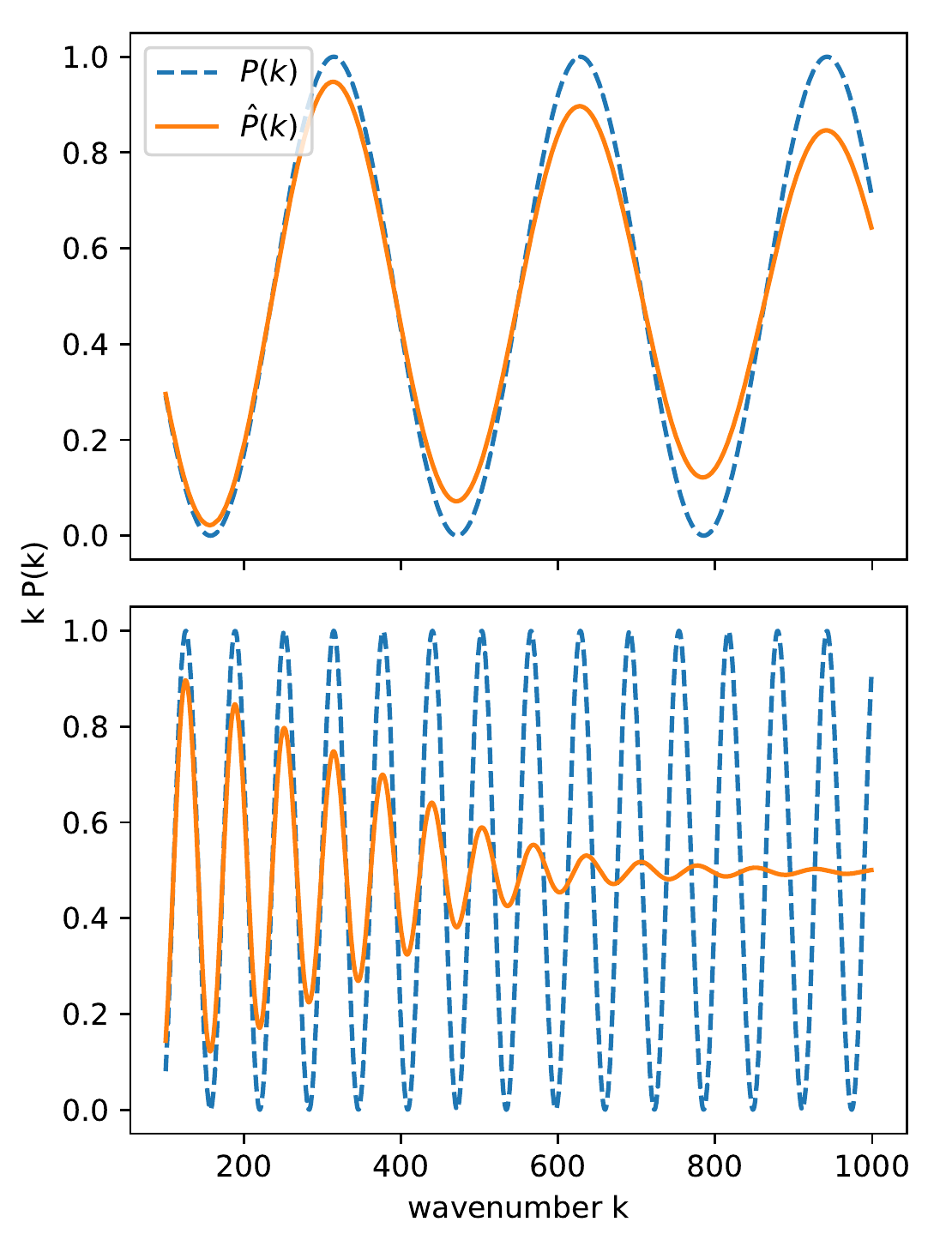}
\caption{An example of the differences between $P(k)$ (thick line) and $\hat P(k)$ (dashed line) for two oscillatory spectra. For illustrative purposes we have set $N=10$ and labelled $k_m$ by $m$. As we see, when the smoothing scale $\Delta k\sim k/N$ is larger than the oscillation scale the oscillations are erased from the sampled spectrum.} \label{fig:oscillations}
\end{figure}

\subsection{The power spectrum for $k\gg k_N$}
The heteroskodastic nature of the distribution, the issues of Gaussianity vs. non-Gaussianity, and the potential breaking of translational invariance become irrelevant in the regime $k\gg k_N$. In this case there are many domains in the field so that the ``end ones'', whose length depends on $x_{0n}$, become sub dominant. The subtleties highlighted in the previous subsection become unimportant and the calculation simplifies. In addition, we can compute the spectrum including leakage terms, and so to a good approximation for realistic power spectra.

If leakage could be neglected, the results found above can be easily explained in this regime. For mode $k$ there are $k/k_N$ domains of size $\ell=N\lambda$. In some domains the mode's amplitude is positive, whereas in others it is negative, so the overall amplitude of the mode is reduced. Its variance 
decreases by a factor of one over the number of realizations/domains over which we are averaging, explaining the suppression by the factor of $k_N/k$ found for $k\gg k_N$. This conclusion would be correct if the power spectrum were a delta function and 
we were observing the single mode with non-vanishing raw power. In practice the spectrum is always reasonably smooth, in which case we find that this dilution effect is almost exactly cancelled by the leakage from other modes into the mode under observation. Indeed the cancellation is exact for a scale-invariant spectrum, as we now show. 

Taking leakage into account, (\ref{hatpk}) can be written in the more compact form:
\begin{equation}\label{newhatpk1}
{\hat P}(k_m)=\sum_{nj}P(k_n)|\tilde W_{n,j}(k_m-k_n)|^2\,,
\end{equation}
(note that the $f^j_{nm}$ introduced in the Appendix are nothing but the Fourier transform of the $W_{n,j}(x)$ window). Thus,
\begin{equation}
{\hat P}(k_m)=\sum_{nj}P(k_n)\left(\frac{\ell_{n,j}}{L}\right)^2
{\rm sinc}^2\left[\frac{(k_n-k_m)
\ell _{n,j}}{2}\right]\,,
\end{equation}
where ${\rm sinc}(x)\equiv\sin x/x$ for $x\neq0$ and ${\rm sinc}(x)=1$ for $x=0$,
and $\ell _{n,j}$ is the size of the domain labelled by $n$ and $j$ (see Appendix).
Let us assume $k_m,k_n\gg k_N$, so that the majority of the domains defined by scale $k$ are contained inside the box. The effect of $x_{n0}$ on the observed spectrum and its leakage is then negligible, since we can ignore the end domains.  The difference between marginal and conditional power spectra is accordingly small.  

Then, $\ell_{n,j}=L N/n$ and the number of domains for mode $n$ is $J_n=n/N$, leading to:
\begin{equation}\label{eq:sum}
{\hat P}(k_m)=\sum_{n}P(k_n) \frac{N}{n}
{\rm sinc}^2 \left[\pi (n- m)\frac{N}{n}\right]\,.
\end{equation}
Since we are working in one dimension (even though the results presented here generalize easily for three dimensions), we have for the dimensionless power spectrum:
\begin{equation}
k P(k) = A^2\left(\frac{k}{k_p}\right)^{n_S-1}\, ,
\end{equation}
with scale-invariance given by $n_S=1$. Approximating the sum by an integral (a good approximation, since $n\gg N$), and setting
\begin{equation}
x=\pi(n-m)\frac{N}{n}\,,
\end{equation}
we find
\begin{equation}
{\hat P}(k_m)\approx \frac{P(k_m)}{\pi}\int^\infty_{-\infty}\,dx\,{\sinc }^2 (x)= P(k_m)\,,
\end{equation}
valid, we stress, for a scale-invariant spectrum only.

Even though there is no sampling effect for a scale-invariant spectrum, there will be a small difference between $\hat P(k)$ and $P(k)$ for any other type of spectrum. Indeed, as can be inferred from the width of the $\sinc$ function in the formulae above, there will be a smoothing of the dimensionless power spectrum on a scale 
\begin{equation}
\Delta k\sim \frac{k}{N}\,.
\end{equation}
This will affect any tilted spectrum, introducing a bias towards scale-invariance, the more so the higher the $k$ and the tilt. This is illustrated in Fig.~\ref{fig:tilted} where we evaluate the sum (\ref{eq:sum}) explicitly for $n > N$. We see that the effect is as expected with red spectra (relevant for the dark matter spectrum on small scales) being biased towards scale-invariance.  The effect would have most impact if there are deviations from a pure power law in the primordial spectrum on these scale. To illustrate this we process an initial spectrum that is oscillatory. In Fig.~\ref{fig:oscillations} we see that the effect is to smooth out the structure in the underlying spectrum. This effect might be relevant baryon acoustic oscillations, but also for the Doppler peaks. However, in order to fully study these, one must study how the domain structure affects evolution, by considering perturbations with non-vanishing momentum, a matter we address in the next Section. 

\section{Unsqueezing and Doppler Peaks' erasure}\label{sec:unsqueeze}
The fact that at late times the fluctuations form {\it standing} 
(rather than {\it travelling}) waves is essential for the existence of
Doppler peaks in the CMB.
The formation of standing waves is usually attributed to ``squeezing''
(although quantum squeezing is not strictly
necessary, a matter investigated in~\cite{squeeze}). Here we show that
the domain structure we have unveiled may ``unsqueeze'' the waves. This happens if, through evolution, a sufficient number of wavelengths have travelled away from their original domain by the time of last scattering. Evolution, therefore, may qualitatively change the conclusions of the last Section, which assumes that the primordial spectrum is frozen in, or that the fluctuations have zero momentum. This is particularly relevant for radiation fluctuations. 

Standing waves are often the result of imposing fixed spatial boundary conditions, but not always. A notable exception is cosmology. A standing wave of a given wavelength can be seen as two travelling waves with the same wavelength moving in opposite directions, constrained to have the same amplitude~\cite{squeeze}. In cosmology, it is the prevalence of a growing mode over a decaying mode while the modes are outside the horizon (i.e. ``squeezed'') that imposes this correlation between modes moving in opposite directions after they re-enter the horizon. As long as the wave train is infinite the wave remains standing, even without imposing fixed boundary conditions.

Specifically, setting $\zeta=v/z$, and writing~\cite{squeeze}:
\be
v(\vx,\eta)=\sum_{\vk} v(\vk,\eta)e^{i\vk\cdot x} + c.c.\,,
\ee
with the sum taken over just one half of $\mathbb{R}^3$, we have the general solution for radiation:
\be\label{eq:expansion}
v(\vk,\eta)=v_0(\vk)e^{-ic_s k\eta}+ v_0^\star (-\vk)e^{ic_s k\eta}\,,
\ee
where the argument in $v_0(\pm \vk)$ denotes the direction of motion of the travelling wave~\footnote{If we ignore baryons, then $c_s=1/\sqrt{3}$, but baryons can easily be included in the discussion.}. After squeezing we have the constraint 
\be\label{constraint}
v_0(\vk)=-v_0^\star(-\vk)\,,
\ee
so that 
\be
v(\vk,\eta)=-2|v_0(\vk)|e^{i\phi_\vk}\sin(c_sk\eta)\,,
\ee
(where $\phi_\vk$ is the phase of $v_0(\vk)$ and determines the positions of the nodes of 
the standing wave). Propagating until last scattering, $\eta=\eta_\star$, and evaluating the power spectrum:
\be\label{Pspecdef}
\langle v(\vk,\eta_\star)  v^\star  (\vk ', \eta_\star)\rangle=\delta(\vk -\vk')P_{v}(k,\eta_\star)\,,
\ee
we find:
\be\label{Ppeak}
P_v(k,\eta_\star)=4P_{v0}\sin^2 (c_sk\eta_\star)\,,
\ee
and thus the oscillations in the CMB power spectrum known somewhat erroneously as Doppler peaks~\footnote{The ``Doppler'' peaks and troughs would be better described as ``density oscillations'', and in fact are softened by the Doppler velocity term, which is out of phase and partly fills the density valleys. The Bessel function projectors between $k$ and angle also smooths the oscillations. We do not need these details in the discussion that follows.}.

Although for an infinite wave train the standing character of the wave remains 
unchanged because (\ref{constraint}) is always valid, this is not true if there is a
domain structure. Then, the two highly correlated
travelling waves making up the standing wave will eventually leave the
domain of origin through opposite sides. The original domain will in turn fill up with
travelling waves coming from adjacent domains in opposite directions. 
Eventually, at any given region, for a given wavelength,
there will still be waves moving in opposite directions, but their
amplitudes will not be correlated, since they come from
uncorrelated domains. Therefore they no longer form standing waves. This only happens for modes which have
had time to move away from their original domains in the time since
they re-entered the horizon and the last scattering surface.  For them, 
the Doppler peaks are erased, as we now explicitly show.

We can quantify a squeezed domain structure by defining: 
\bea
\langle v_{0i}(\vk)  v_{0 j}^\star (\vk ')\rangle&=&\delta_{ij}\delta(\vk -\vk')P_{v0}(k)\label{eq:correlator1}\,,\\
\langle v_{0i}(\vk)  v_{0 j} (\vk ')\rangle&=&\delta_{ij}\delta(\vk +\vk')Q_{v0}(k),\label{eq:correlator2}\,,
\eea
and setting $Q_{v0}(k)=-P_{v0}(k)$ so as to impose constraint (\ref{constraint}) where appropriate (see~\cite{squeeze1} for more detail). 
The labels $i$ and $j$ refer to original domains where the waves were imprinted, where perfect squeezing was present. However, then the waves
move away from each other, leading to unsqueezing if they fully leave the original domains. At time $\eta=\eta_\star$ the waves have travelled $c_s\eta_\star$, to be compared with the size of their domains, $N\lambda$. Hence the outcome of this process (unsqueezing or not) depends on the value of $k$ with respect to:
\be
k_u=\frac{N\pi}{c_s\eta_\star}\,,
\ee
corresponding to the wavelength that has travelled half the domain size
at $\eta=\eta_\star$. This mode has travelled enough to just 
miss any overlap between the two correlated travelling waves that made up the original standing wave. For $k\ll k_u$ the overlap is considerable and the calculation leading to~(\ref{Ppeak}) is approximately valid. For $k\ge k_u$ we have instead that wherever we look:
\be
v(\vk,\eta)=v_{0i}(\vk)e^{-ic_s k\eta}+ v_{0j}^\star (-\vk)e^{ic_s k\eta}\,,
\ee
with $i\neq j$ (all the travelling waves at a given point originated form different domains). Inserting (\ref{eq:expansion}) into (\ref{Pspecdef}) and using (\ref{eq:correlator1}) and (\ref{eq:correlator2}) we thus obtain:
\begin{eqnarray}\label{eq:Pnopeak}
P_v(k,\eta_\star)&=&2P_{v0}(k)(1-2\delta_{ij}\cos (2c_s k\eta_\star)) \nonumber\\
&=&2 P_{v0}(k)\,,
\end{eqnarray}
to be contrasted with (\ref{Ppeak}) (which would be recovered if we could set $i=j$). The Doppler peaks have been erased, as expected, since the waves no longer are standing waves. 

How this might be seen in real life is the subject of the rest of this Section. 
Given that the distance travelled by the waves at last
scattering ($\sim c_s\eta_\star$) is of the order of the wavelength
corresponding to the first Doppler peak, and given that $N$ is large, we 
expect the scale of unsqueezing, $k_u$, to correspond to very small angles. 
In practice, we expect a gradual softening of the
peaks from almost nothing at the first peak, to almost total beyond the $N$th
peak. Such very high resolution is likely to be accessible only by 
interferometry, for which the fields are very small. We will then only sample one domain. This simplifies the calculations enormously 
(Section~\ref{exp1}). Should this not be the case the calculation is more 
elaborate, and a heuristic derivation in presented
in Section~\ref{exp2}. A partial combination with the results in Section~\ref{sec:power} is then needed. 

\begin{figure}[t]
\centering
\includegraphics[trim={6.3cm 1cm 1cm 2cm},clip, width=9.5cm,angle=0]{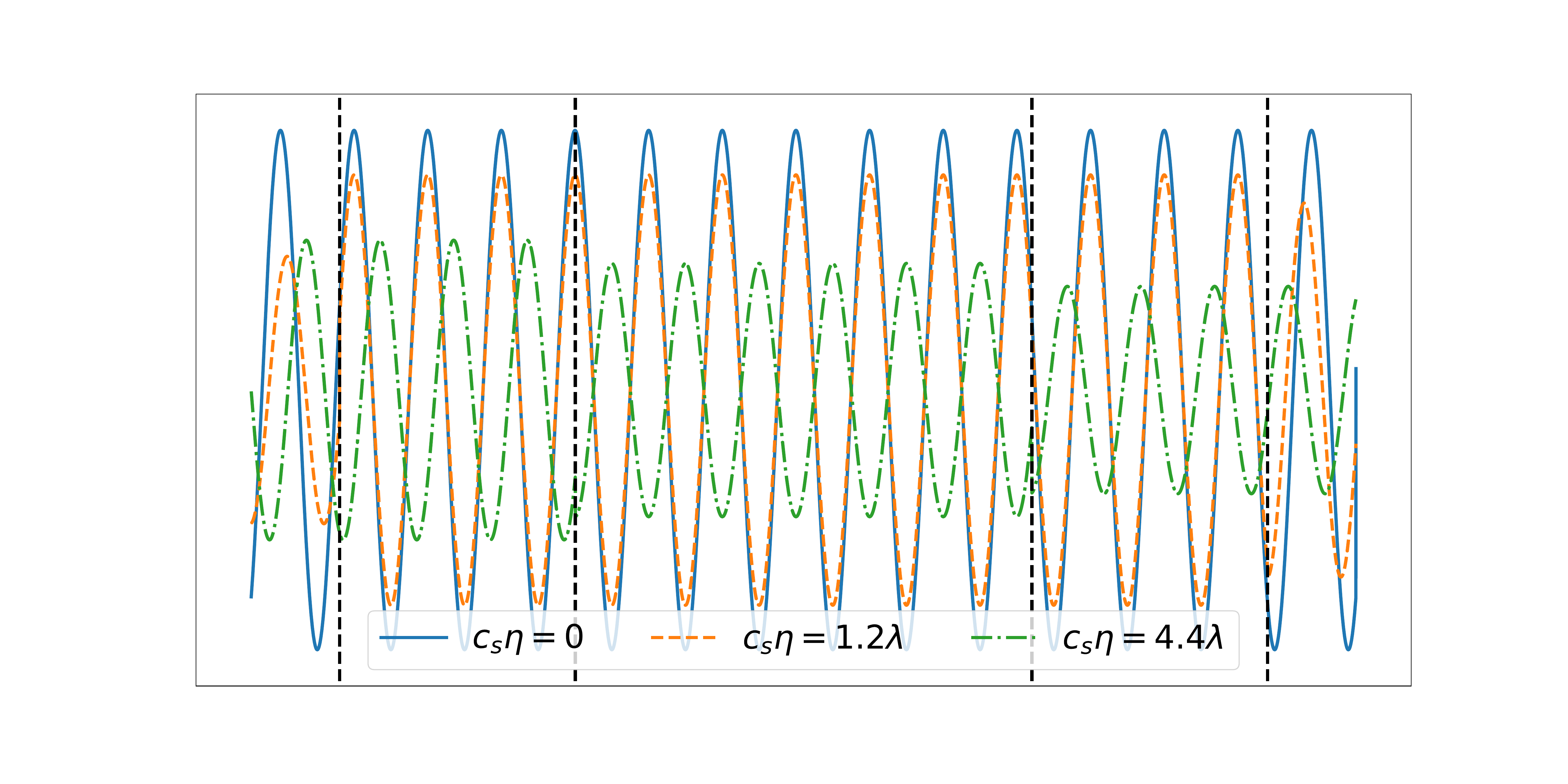}
\caption{An initial domain of size $L=N\lambda$ with correlated  right and left moving waves forms a standing wave at $\eta=0$ (solid, blue). However, as time passes and the waves move past each other, introducing uncorrelated domains from either end, the region where the wave is standing shrinks. Thus at a late time $c_s\eta=4.4 \lambda$ only the central region displays stationary nodes in the wave. In this example $N=15$.}\label{fig:sound_wave}
\end{figure}

\subsection{The measured power spectrum in a simplified case}\label{exp1}
Let us now assume that we probe very small scales with very small fields, so that there is never a significant chance of catching more than one region either with different power spectra (because the travelling waves have moved apart in some regions but not others) or different realizations of the same process (i.e. different domains). This requires the field size, $L$, to satisfy:
\be
L\ll \min\left(c_s\eta_\star,  \frac{N\lambda}{2}\right)\,.
\ee
Then, for $k<k_u$ each domain (with size $N\lambda$) has two sub-regions (on either extreme), each of size $c_s\eta_\star$, where there are no standing waves, since the correlated travelling waves have moved away from each other. The rest of the region (the internal part) is still filled with standing waves, because the correlated travelling waves are still superposed there. This situation is shown in Fig.~\ref{fig:sound_wave} for an example mode and a small value of $N$ for illustrative purposes. If we probe the power spectrum with a field much smaller than $2c_s\eta_\star$, we therefore will only probe a single realization/power spectrum. 
We have a heteroskodastic process, with probability 
\be
{\cal P}=\frac{2c_s\eta_\star}{N\lambda}=\frac{k}{k_u}\,,
\ee
 of measuring power spectrum (\ref{eq:Pnopeak}), for which the peaks have been erased, and the complementary probability of measuring (\ref{Ppeak}), i.e. observing the usual peaks. These are the conditional power spectra, and the  marginalized power spectrum is therefore:
\be\label{margP}
P_v(k,\eta_\star)=2P_{v0}{\left[ \frac{k}{k_u}+2{\left(1-\frac{k}{k_u} \right)}\sin^2 (c_sk\eta_\star)\right]}\,.
\ee

For $k\ge k_u$ only the power spectrum (\ref{eq:Pnopeak}) is observed, at least if only one domain is present in the field. This requires $L\ll N\lambda /2$, but the conclusion should remain valid (in light of previous results) even with domains with  $L<c_c\eta_\star$, since no correlated waves moving in opposite directions will then be present inside the observing region. 

\subsection{More than one domain per field}\label{exp2}
For the case where the observing field is large there may be two possibilities. Firstly, for $L>c_s\eta_\star$ and $k<k_u$, it may be  possible to measure the consequence (\ref{margP}) since in this case we are marginalizing over the domain locations. 

Secondly, for $L>N\lambda/2$ and $k>k_u$. This corresponds to the regime where the decorrelation of the left and right moving waves is maximised and standing waves disappear. However, for relevant scenarios, this corresponds to scales where the dilution from the probabilities (\ref{eq:Pnopeak}) and (\ref{Ppeak}) also determines the observed spectrum. Fundamentally, we are limited by the fact that we can only observe a number of independent modes at a given wavelength and that recombination happened at early times ensuring that $c_s\eta_\star \ll N\lambda$ and that $k_u \to k_N$. This means that any observable effect would be at small scales, close to or smaller than $N$ times the scale of the first acoustic peak. It is difficult to imagine how these scales would be probed in the CMB for values of $N\sim 1000$ motivated in our simple scenario. The situation would be very different if recombination had happened much later. In that case we would be able to observe the decorrelation of acoustic peaks even at larger angular scales. 

A rigorous calculation of the expected effects of this scenario on the CMB are beyond the scope of this work. The simple one dimensional heuristic picture we have introduced could be extended to three dimensions by considering domains arranged along the line-of-sight. This would reconcile the setup with isotropy. It is also important to note that significant convolution of three dimensional Fourier domain modes occurs due to the projection of plane waves onto the last scattering surface, although this is more important on large scales where the effects of this model are less important as discussed above.

\section{Unsqueezing of the Gravitational Wave Background}\label{sec:gws}

Could unsqueezing be observed directly by determining if a background
is made up of travelling or standing waves? A coherent
measurement from a single location is not sufficient to distinguish
between the two possibilities but a correlation of signals from two
coherent measurements at different locations is. Indeed the fact that a radio interferometer measures coherent travelling waves leads to a time dependent signal that is usually averaged out to recover the intensity of the signal. 

Gravitational wave interferometers can be considered as analogues to
radio interferometers except that each detector is equally sensitive
to waves moving in opposite directions - the beam is simply a function of the angle and polarisation of the incoming wave with respect to the single detector baseline. To see how this may be used to
determine the nature of the waves we can expand a continuous spectrum
of gravitational waves \footnote{We disregard the polarization of the waves in this discussion.} at a location $\vx$ and time $\eta$ as
\begin{equation}
  h(\vx,\eta) =\int_0^\infty  
  \nu^2 d\nu d\Omega \, h(\nu,\eta)\, e^{i 2\pi
  \nu\vhk\cdot \vx} + c.c.\,,
\end{equation}
where 
$\nu$ is the frequency of the incoming wave with line-of-sight $\vhk$, and $d\Omega$ is
the infinitesimal area element on the sky. The mode functions
$h(\nu,\eta)$ are expanded as in (\ref{eq:expansion})
\begin{equation}
h(\nu,\eta)=h^{\,}_{0i}(\nu)e^{-i2\pi\nu \eta}+ h_{0j}^\star (-\nu)e^{i
  2\pi\nu\eta}\,,
\end{equation}
but we have assumed a standard dispersion relation to emphasise the frequency domain nature of the measurement. The case of a background made up of standing waves is recovered if the domain size approaches infinity, in which case $i=j$ at all times with amplitudes and time phase correlated between and left and right moving components. Otherwise, for a finite domain size, the superposition of uncorrelated left and right components will result in a travelling wave with non vanishing momentum. 

\begin{figure}[t]
\centering
\includegraphics[width=9cm,angle=0]{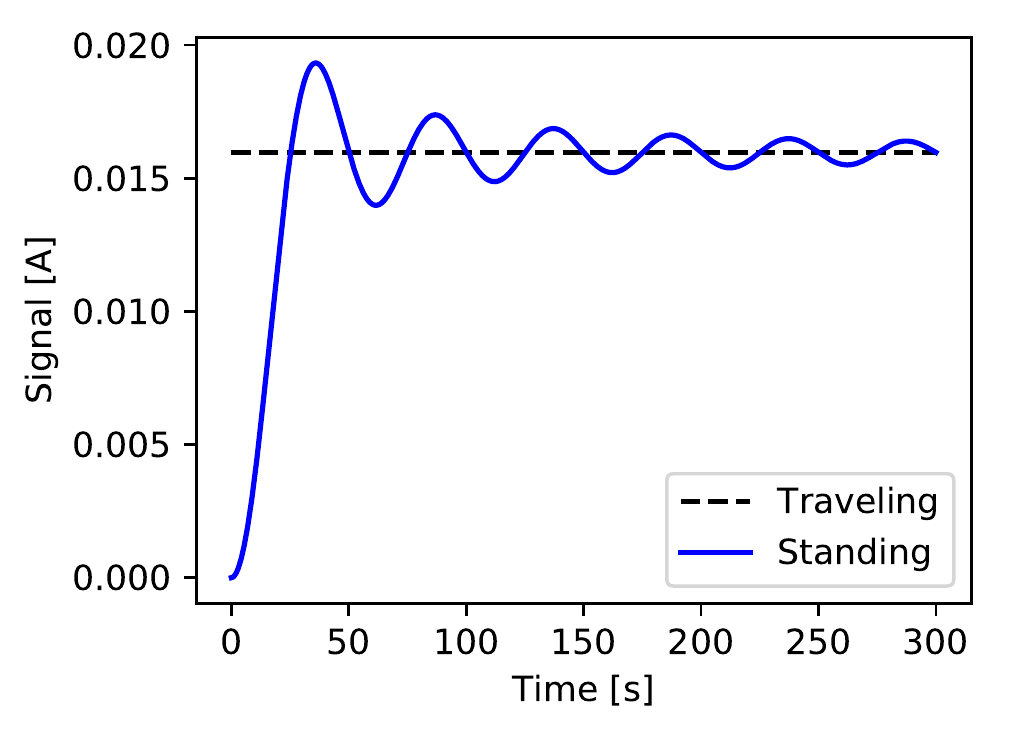}
\caption{The interference signal produced by travelling and standing waves, with a constant spectrum of amplitude $A$, integrated uniformly over the whole sky for two detectors separated by $2.5\times 10^6$ km and with uniform sensitivity to frequencies up to 0.01 Hz. Note that the domain crossing time for the highest frequency would be 100,000 s for $N=1000$. }\label{fig:analytic}
\end{figure}

We can now apply the formalism developed above and in \cite{squeeze,squeeze1} to the cross-correlation of detectors separated by
a baseline vector $\vb=\vx-\vy$. The ensemble average of the product 
of the signals seen at each detector is $s(\eta) = \vev{h(\vx,\eta)h(\vy,\eta)}$. We can expand this in terms of the correlators
(\ref{eq:correlator1}) and (\ref{eq:correlator2}) to obtain 
\begin{eqnarray}
s(\eta) &=& \frac{2}{\pi^2} \int_0^\infty  \nu^2
  d\nu d\Omega \,\cos\left(2\pi\, \nu |\vb|\, \vhk
\cdot\vhb\right)\times \nonumber\\
&&\left[2P(\nu)+Q(\nu)e^{-i
  4\pi\nu\eta}+Q^\star(\nu)e^{i
  4\pi\nu\eta}\right]\,.
\end{eqnarray}
In the case of a travelling wave the spectrum $Q(\nu)=0$ and as such the time oscillating components of the integrand above disappear. In the standing wave case we have $Q(\nu)=Q^\star(\nu)=-P(\nu)$
we therefore have and integration over a set of plane wave in time. Carrying out the integral over all orientations of the line-of-sight and  \footnote{A realistic calculation would insert the correct response radiometer response function at this stage but we omit it in this simple exercise.} we obtain and expression for the time-dependent correlator
\begin{equation}
s(t) = \frac{8}{\pi^2} \int_0^\infty  \nu^2 d\nu
  \sinc\left(2\pi\, \nu |\vb|\right) P(\nu)\left[1-\cos(
  4\pi\nu\eta)\right]\,.
\end{equation}
The distinguishing feature in the above is the presence of a cosine modulation of the frequency integrand in time which is absent for the case $Q(\nu)=0$.

In Fig.~\ref{fig:analytic} we show the expected signal for travelling and standing waves with a constant spectrum $P(\nu)=A$ and for an instrument having a uniform frequency response up to a maximum $\nu_{\rm max} =0.01$ Hz and a baseline length of $2.5\times 10^6$ km. The parameters are chosen as typical ones of a space--based gravitational wave interferometer such as LISA \cite{LISA}. We see that the coherence of standing waves induces an oscillation which is dominated by a mode that is twice the maximum frequency included in the signal. The addition of a high-pass cut-off in the signal would induce a dominant long time--scale mode on top of that seen in the figure.

Clearly, sample and noise variance would affect the ability to measure these damped oscillations  characteristic of a standing wave background. We envisage the ensemble average limit being approximated by averaging over repeated measurements of the signal over times--scales much smaller than the domain crossing time, $N/\nu_{\rm max}$, of the highest frequency mode the instrument is sensitive to.

A comment on the wider context of gravitational wave background observations is in order here. Our distinction between the interferometric signal of travelling and standing waves is of use to determine the nature of {\it any} background measurement. For example, a stochastic background made up of the signal of overlapping, unresolved sources will not display the coherent oscillations of a standing wave. Therefore, our proposed measurement is of use in distinguishing stochastic backgrounds from any coherent cosmological ones. 

\section{Discussion}\label{sec:discussion}

In cosmology, the coherence of fluctuations about the background on super-horizon scales is a fundamental principle in  our current understanding of the subject. It is supported by clear evidence in the form of multiple acoustic peaks in the CMB. However, we are used to extrapolating this assumption to much smaller scales where there is both a lack of observational verification {\it and} potential new physics affecting the nature of causality. In this paper we have discussed how introducing a simple model where coherence is limited spatially on a characteristic scale during inflation can lead to testable predictions.

We have developed the formalism to deal with the heteroskodastic nature of realisations and the loss of spatial phase coherence in such a scenario. We have shown how describing the level of coherence of background fluctuations in terms of temporal and spatial phase correlations of waves moving in opposite directions helps to understand whether a background consist of travelling or standing waves and how this interpretation can depend on the scale considered in this scenario. 

To determine observational consequences in this model we have looked at two broad categories of fluctuations. The first is a background of density fluctuations with no momentum, such as a cold dark matter fluid. In this case the presence of finite domains representing different realizations results in subtle consequences for the observed spectrum of perturbations. On wavelengths larger than the characteristic scale set by the domain size, the effect is to leave the power spectrum of the fluctuations unchanged. On scales close to the characteristic scale however, the heteroskodasticity of the realization leads to important consequences. For a given realization of domain centres there are off-diagonal correlations of the modes due to coupling induced by the windowing of the Gaussian realisation. However, if our observations effectively marginalize over the entire distribution of domain centres then the off-diagonal correlations disappear but the realisation becomes non-Gaussian with non-vanishing {\it higher-order} cumulants. 

On scales much smaller than the characteristic scale we find that the dominant effect is due to the leakage, or coupling of modes, induced by the large number of domains inside the observational window. This leads to an effective convolution of the power spectrum with the shape of the domain, which biases the spectrum towards scale invariance with respect to the underlying spectral tilt and would tend to erase features in the underlying power spectrum. Only for a scale-invariant power spectrum would we see no effects. 

The second category of backgrounds is that made up of waves with non-vanishing momentum such as fluctuations in the photon-baryon fluid before recombination observed in the CMB or a cosmological background of primordial gravitational waves. In the CMB case, the onset of uncorrelated {\it travelling} waves with opposite momentum, formed as waves from coherent domains (initially correlated so as to form {\it standing} waves) stream away from each other, will erase the presence of acoustic peaks. This effect is important on small scales which translate to multipoles on the sky $\ell_u\sim N\ell_\star$ where $\ell_\star\sim 220$ is the scale of the first acoustic peak. Thus the effect would be very difficult to measure if $N\sim 1000$ (but, of course, we could envisage scenarios where $N$ is smaller). We note that there may be a more realistic opportunity to measure this effect in the baryon acoustic signature of large scale structure observations and we leave this for future work. 

The situation is very different for the gravitational wave case, since we probe these waves directly on scales that are much smaller than the expected domain size. In this case we have shown that there is a clear distinction in the time coherence of a signal measured by interferometry between a background made up of travelling or standing waves. In principle, the correlation of two gravitational wave detector signals contains a short time--scale modulation that is a characteristic signal of a standing wave background. This is the same signal present in radio interferometry when correlating two coherent detectors but in that case it is present when observing {\it travelling} waves from a {\it single} direction. Further study is required to determine how difficult it would be to measure this signal. It should also be noted that we expect there to be multiple stochastic backgrounds of gravitational waves, mostly due to the confused, incoherent superposition of the emission of astrophysical objects. These backgrounds will be made up of travelling waves. They are also expected to be larger than most predicted primordial signals such as that produced during a period of inflation in the early universe. An important challenge for future gravitational wave detectors will be the separation of these stochastic backgrounds in the search of a primordial signal. The signature of standing wave backgrounds we have introduced here may play a role in addressing this challenge.

We have carried out an initial investigation of this type of model using heuristic arguments involving one dimensional realisations in most cases. It would be interesting to develop more generalised calculations of the effects on two fronts. Firstly, there is much scope to explore the the statistics and theory of estimation  in cases involving higher--dimensional heteroskedastic correlated fields. Secondly, it may be worthwhile exploring the decoherence of fluctuations in coordinate space as opposed to the Fourier domain approach taken in this work. For example, even the small amount of decoherence we have argued may occur on acoustic peak scales may lead to stronger effects in coordinate space statistics, such as those involving peak counts and topological measures. Mixed Fourier/coordinate space methods such as wavelet analysis may also be useful in analysing the consequences of these models, although we note that our calculations involving windowed Fourier transforms are themselves a form of wavelet analysis.

\begin{acknowledgments}
The authors would like to thank Marco Peloso for discussions in the initial stages of this work. We also thank Robert Brandenberger and Giulia Gubitosi for discussions. JM was sponsored by the John Templeton Foundation during part of this project. Both authors were funded by an STFC consolidated grants ST/L00044X/1 and ST/P000762/1. 
\end{acknowledgments}

\appendix

\section{Evaluation of the power spectrum of a system of independent domains}\label{app:domains}

We consider the Fourier transform of a finite box that is sub-divided into uncorrelated realizations of the same underlying spectrum (the ``domains''). Since the domain size is a function of the wavelength being considered, there will be a scale $k_N$ corresponding to the domain size that is equal to the box size $L$. As explained in the main text, this crucial scale is:
\begin{equation}
k_N=\frac{2\pi}{L}N.
\end{equation}
We start by evaluating the number of domains in the field for a given $k_n$.

\subsection{Number of domains and their location}
For a given $k_n=\frac{2\pi}{L}n$
we introduce $J_n$, the number of domains present in the box of size $L$ for that wavelength. This depends on $x_{0n}$, and since each domain has size $l_n=L N/n$, we must have $-l_n<x_{0n}<0$ to avoid redundancy. For clarity, we start by separating the 
two cases $n\le N$ ($k_n<k_N$) and $n>N$ ($k_n>k_N$). If $n\le N$, then $l_n\ge L$, and we have 
\begin{itemize}
\item 
$J_n=1$ if:
\begin{equation}
-(l_n-L)<x_{0n}<0\,,
\end{equation}
\item $J_n=2$ if:
\begin{equation}
-l_n <x_{0n}<-(l_n-L)\,.
\end{equation}
\end{itemize}
In the first case we have a single domain in our box; in the second case we have two domains, with a change at $x_{n,2}$. If $n> N$ we have 
\begin{itemize}
\item 
$J_n=\left[ \frac{n}{N} \right] + 1$ if:
\begin{equation}
-\left(L-\left[ \frac{n}{N} \right] l_n\right)<x_{0n}<0\,,
\end{equation}
\item $J_n=\left[ \frac{n}{N} \right] + 2$ if:
\begin{equation}
-l_n<x_{0n}<-\left(L-\left[ \frac{n}{N} \right] l_n\right)\,.
\end{equation}
\end{itemize}
In the first case we have a first partial domain of size $l_{n1}=l_n+x_{0n}$, then $\left[ \frac{n}{N} \right]-1$ full domains of size $l_{nj}=l_n$, and finally a partial domain of size $l_{n J_n}=L-l_n\left[ \frac{n}{N} \right]-x_{0n}$. In the second case we have a first partial domain of size $l_{n1}=l_n+x_{0n}$, then $\left[ \frac{n}{N} \right]$ full domains of size $l_{nj}=l_n$, and finally a partial domain of size $l_{n J_N}=L-l_n\left[ \frac{n}{N} \right]-(x_{0n}+l_n)$. Notice that our formula for $n>N$ actually contains the one for $n\le N$. We have separated the two cases just to make the discussion clearer. 

In all cases the $x_{0n}$ are independent and uniformly distributed in their domain ($-l<x_{0n}<0$), with probability:
\begin{equation}
{\cal P}(x_{0n})=\frac{1}{l_n}.
\end{equation}

\subsection{The processed power spectrum}
In order to compute the processed power spectrum given the domain structure and a given observation box of size $L$, we now Fourier Transform the random field $\zeta$ using 
\begin{equation}
\zeta \left( x \right) = \sum_{m=1}^\infty {\hat \zeta} \left( k_m \right) {\rm e}^{i k_m x}\,,\end{equation}
and
\begin{equation}
 {\hat \zeta} \left( k_m \right)  = \frac{1}{L} \int_0^L d x \, \zeta \left( x \right) \, {\rm e}^{-i k_m x}\,.
\end{equation} 
Then,
\begin{eqnarray}\label{FT-exp}
 {\hat \zeta} \left( k_m \right)  &=& \sum_{n=1}^\infty \sum_{j=1}^{J_n} \zeta_{n,j} \, \frac{1}{L} \int_0^L d x \, {\rm e}^{i \left( k_n - k_m \right) x} \, W_{n,j} \left( x \right) d x\,, \nonumber\\
 &\equiv&  \sum_{n=1}^\infty \sum_{j=1}^{J_n} \zeta_{n,j} \, f_{nm}^j\, .
 \end{eqnarray} 
The $f$ elements in general have the form:
\begin{eqnarray} 
f_{nm}^j &=& \frac{1}{L} \int_{\max(0, x_{n,j})}^{\min(L, x_{n,j+1})} d x \, {\rm e}^{i \left( k_n - k_m \right) x} \,,\nonumber\\ 
&=& \delta_{nm} \, \frac{l_{nj}}{L} + \delta_{n \neq m} \,g^j_{mn}\,,
\end{eqnarray} 
where the second term is a leakage term. If the domain $j$ is fully contained inside the box this can be evaluated as:
\begin{eqnarray}\label{fnm}
f_{nm}^j&=& \delta_{nm} \, \frac{N}{n} +  \\
&&\delta_{n \neq m}
\, {\rm e}^{2 \pi i \left( n - m \right) \left(j \frac{N}{n} + \frac{x_{0n}}{L}\right) } \, \frac{1-{\rm e}^{-2 \pi i \left( n - m \right) \frac{N}{n}}}{2 \pi i \left( n - m \right)} \,.\nonumber
\end{eqnarray}
Otherwise the second term is more complicated. 
We seek to compute the processed power spectrum given the domain structure:
\begin{equation} 
\left\langle \hat \zeta_{m} \hat \zeta^*_{m'} \right\rangle = \delta_{m m' } \, \hat P \left( k_m \right) .
\end{equation} 
From (\ref{corr-ni}) and (\ref{FT-exp}) we see that the sampled power spectrum is:
\begin{eqnarray}\label{hatpk}
{\hat P}(k_m)&=&\sum_{nj}P(k_n)|f^j_{nm}|^2\,, \\
&&=\sum_j\frac{l_j^2}{L^2}P(k_m)+\sum_{nj}P(k_n)|g^j_{nm}|^2\,,\nonumber
\end{eqnarray}
so that we see the second, off-diagonal term in $f^i_{nm}$ is a leakage term. 
We will compute both the power spectrum conditional to a set of $\{x_{0n}\}$, as well as a power spectrum marginalized over these
\begin{equation}
\tilde P=\int dx_{n0} {\cal P}(x_{n0}) \hat P(k |x_{n0}). 
\end{equation}

\subsection{The processed power spectrum ignoring leakage}
We now evaluate the processed power spectra (conditional and marginal) ignoring leakage. This will clarify a number of issues in the transition zone between the two extreme regimes. Unfortunately this is a bad approximation, except in the case where the raw spectrum is a delta function for a given mode and then we focus on that specific mode. We shall carry out this exercise in this subsection.

\subsubsection{Power spectrum for $n\le N$ ($k\le k_N$)}
If $n\le N$, the domain size $l_n=LN/n$ is larger than $L$ and depending on $x_{0n}$ we can either have 1 or 2 domains inside the box.
If $-(l_n-L)<x_{0n}<0$, we have a single domain, and from (\ref{hatpk}), ignoring 
the leakage term, we get:
\begin{equation}\label{Pkcond1}
\hat P(k|x_{n0}) = P(k)\,,
\end{equation}
as expected. The probability for this to happen is 
\begin{equation}
{\cal P}\{-(l_n-L)<x_{0n}<0\}=\frac{l_n-L}{l_n}=1-\frac{n}{N}\,.
\end{equation}
If $-l_n <x_{0n}<-(l_n-L)$, there is a change of domain at $x_{n,2}=x_{n0}+l$ and
\begin{equation}\label{two_domain}
\hat P(k|x_{n0})= P(k)\frac{x_{n,2}^2+(L-x_{n,2})^2}{L^2}\,. 
\end{equation}
The processed power is therefore reduced, and the effect is maximal (reduction by 1/2) if the change of domain occurs at the mid point of the observation box. The probability for a change of domain occurring at $x_{n,2}$ is:
\begin{equation}
{\cal P}(x_{n,2})=\frac{n}{N}\frac{1}{L}\,,
\end{equation}
i.e. it is uniformly distributed in $[0,L]$ but does not integrate to one, to account for the fact that a change of domain may not occur. 

We can now compute the marginalized power spectrum as:
\begin{equation}
\tilde P=\int dx_{n0} {\cal P}(x_{n0}) \hat P(k, |x_{n0})=P(k)\left(1-\frac{1}{3}\frac{n}{N}\right)\,,
\end{equation}
and since $n/N=k_n/k_N$ the final result can be presented as:
\begin{equation}
\tilde P(k)=P(k)\left(1-\frac{1}{3}\frac{k}{k_N}\right)\,.
\end{equation}

\subsubsection{Power spectrum for $n> N$ ($k>k_N$)}

If $k\gg k_N$ the discussion simplifies, but only asymptotically,  not in the transition region. 
We first compute the power spectrum conditional to a given $x_{0n}$. 
If $ -\left(L-\left[ \frac{n}{N} \right] l_n\right)<x_{0n}<0$, taking into account the various $l_{nj}$ computed above, from (\ref{hatpk}), ignoring the leakage term, we get:
\begin{eqnarray}
\hat P(k|x_{n0})&=& P(k)\left\{\frac{l_{n1}^2}{L^2}+ \frac{l_{nJ_n}^2}{L^2}+
\left(\left[ \frac{n}{N} \right] - 1\right)\left(\frac{N}{n}\right)^2\right\}\,,
\nonumber\\
l_{n1}&=&x_{0n}+l_n\,,\nonumber\\
l_{nJ_N}&=&L=l_n\,.
\end{eqnarray}\label{Pkcond3}

If $-l_n<x_{0n}<-\left(L-\left[ \frac{n}{N} \right] l_n\right)$ we have
\begin{eqnarray}
\hat P(k|x_{n0})&=& P(k)\left\{\frac{l_{n1}^2}{L^2}+ \frac{l_{nJ_n}^2}{L^2}+
\left[ \frac{n}{N} \right] \left(\frac{N}{n}\right)^2\right\}\,,
\nonumber\\
l_{n1}&=&x_{0n}+l_n\,,\nonumber\\
l_{nJ_N}&=&L-l_n\left[ \frac{n}{N} \right]-(x_{0n}+l_n)\,,
\end{eqnarray}\label{Pkconf4}
with similar expressions in terms of $k$ and $k_N$ noting that $l_n/L=N/n=k_N/k$.
The marginalized power spectrum is
\begin{equation}
\tilde P(k)= P(k)\left[\frac{k_N}{k}-\frac{1}{3}\left(\frac{k_N}{k}\right)^2\right]\,.
\end{equation}
The above discussion illustrates the heteroskodastic nature of the distribution in the transition regime, as discussed in the main text of the paper.

\end{document}